# Surface Plasmon Assisted Control of Hot-Electron Relaxation Time

Sarvenaz Memarzadeh[1,2,†], Jongbum Kim[2,3,†], Yigit Aytac[4],

Thomas E. Murphy[1,2], Jeremy N. Munday[1,2,3*]

[1]Department of Electrical and Computer Engineering, [2]Institute for Research in Electronics and Applied Physics, University of Maryland, College Park, Maryland 20742, United States

[3]Department of Electrical and Computer Engineering, University of California, Davis, California 95616, United States

[4]Science Systems and Applications, Inc. Lanham, Maryland 20706, United States

**Abstract**

Surface plasmon mediated hot carrier generation is widely utilized for the manipulation of the electron-photon interactions in many types of optoelectronic devices including solar cells, photodiodes, and optical modulators. A diversity of plasmonic systems such as nanoparticles, resonators, and waveguides have been introduced to enhance hot carrier generation; however, the impact of the propagating surface plasmons on hot carrier lifetime has not been clearly demonstrated. Here, we systematically study the hot carrier relaxation in thin film gold (Au) samples under surface plasmon coupling with the Kretschmann configuration. We observe that the locally confined electric field at the surface of the metal significantly affects the hot carrier distribution and electron temperature, which results in a slowing of the hot electrons relaxation time, regardless of the average value of the absorbed power in the Au thin film. This result could be extended to other plasmonic nanostructures, enabling the control of hot carrier lifetimes throughout the optical frequency range.

**Introduction**

Recently, the optical generation of hot carriers in metallic components has attracted interest for applications such as solar energy conversion[1–5], non-linear optics[6–8], sensitive photodetectors[9–12], nanoscale heat sources[13], photochemical reactions in biomolecular studies[14–16], and biosensors[15,17,18]. For the excitation of hot carriers in metals, the incident photon energy is typically lower than the energy of the band-to-band transition, thus the efficiency of hot carrier generation is reduced as a result of the poor absorption of light within the metals. To overcome this limitation, surface plasmons have been broadly utilized to enhance absorption through the use of metallic nanostructures[19–23], which increase the measurement sensitivity because of the increased absorption[24]. Furthermore, the Epsilon-Near-Zero (ENZ) mode in metallic semiconductors has also recently been used to improve the photon-electron interaction for the enhancement of hot carrier generation[25,26].

Hot carriers relax to equilibrium through plasmon dephasing via Landau damping, electron-electron (e-e) scattering, electron-phonon (e-ph) scattering, and lattice heat dissipation through phonon-phonon (ph-ph) interactions[27]. Throughout these processes, hot carriers can distribute their energy to the surrounding environment and in turn thermalize from their excited state to equilibrium. The temporal duration of hot carrier relaxation is the key factor to determine the performance of hot carrier devices. For example, the efficiency of hot carrier injection in energy conversion systems[1–3,5,28] and the operating speed in optical modulation systems[29,30] are both strongly linked to hot carriers' lifetime. Depending on the geometry of metal nanostructures, the materials' band structure, and the incident photons' energy[31], the relaxation time can vary from a few hundred femtoseconds up to a couple of picoseconds. In the case of gold and aluminum nanostructures, relaxation times on the order of hundreds of picoseconds, due to the acoustic vibrations of the lattice, have been reported[32–34]. The effect of enhanced absorption on hot carrier relaxation time has been extensively studied in the case of te thin film and nano-structured plasmonic systems [35–37]; however, the importance of the strongly confined field inside the metal thin film induced by surface plasmon coupling on hot-carrier lifetimes is still elusive.

Transient reflectivity measurements using pump-probe spectroscopy are a common method to characterize carrier dynamics under the intra-band or inter-band transitions. Typically, the measured transient signals for pump-probe spectroscopy are analyzed with the Two-Temperature Model (TTM), which describes the spatiotemporal profile of the electron and the lattice temperature from a coupled nonlinear partial differential equation[38–40]. This model is very useful in understanding relaxation dynamics, but appropriate modification is needed for an accurate modeling of the unique internal electric field profile in metal films due to its coupling to the propagating surface plasmon.

In this work, we investigate the relationship between the hot carrier relaxation time and the characteristics of surface plasmon coupled on the surface of gold (Au) thin film under the Kretschmann configuration. For accurate theoretical modeling of the transient reflectivity data resulting from the carrier dynamics in the conduction band of Au thin film, we employ the free electron model to estimate the elevated electron temperature due to intra-band optical pumping. From the calculated electron temperature, we extract the carrier relaxation time with the modified TTM to better describe the localized electric field distribution inside the Au thin film. Under fixed absorbed power in the Au film over the spectral range of 730 nm to 775 nm (resonance wavelength at 745 nm), we observe that the hot-electron relaxation time in Au film reaches its maximum at the resonance wavelength, which indicates that the modified intensity and profile of the internal electric field by the excitation of surface plasmons plays the significant role in hot carrier relaxation.

**Results and discussions**

To study the effect of surface plasmons on hot-electron relaxation dynamics, we combine the prism coupling technique under the Kretschmann configuration, schematically illustrated in Fig. 1a, with standard degenerate pump-probe optical spectroscopy. Further details on the optical set-up are described under Methods. The thickness of the Au film and the incident angle of light are set to 44 nm and 44°, respectively. Under these conditions, surface plasmon excitation occurs at 745 nm (1.66 eV), where the photon energy is lower than the d-band transition of Au, at 2.4 eV[41]. Once the surface plasmon is excited in the Au film, the electric field is strongly confined at the interface between the Au film and air. Figure 1b shows absorption as a function of wavelength ranging from 730 nm to 775 nm, with resonance wavelength at 745 nm (see Fig. S1 for the broad range of absorption spectrum).

To rule out the effect of absorbed light power in the control of the hot carrier relaxation temporal dynamics, we designed two different experimental conditions: 1) sweeping the wavelength ($\lambda = 730 \sim 775$ nm) with the fixed absorbed power ($P_{abs} = 120 \ mW$), and 2) varying the absorbed laser power ($P_{abs} = 50 \sim 150 \ mW$) with the fixed wavelength ($\lambda = 745$ nm). Figure 1c and 1d schematically illustrate hot electron excitation under these two conditions.

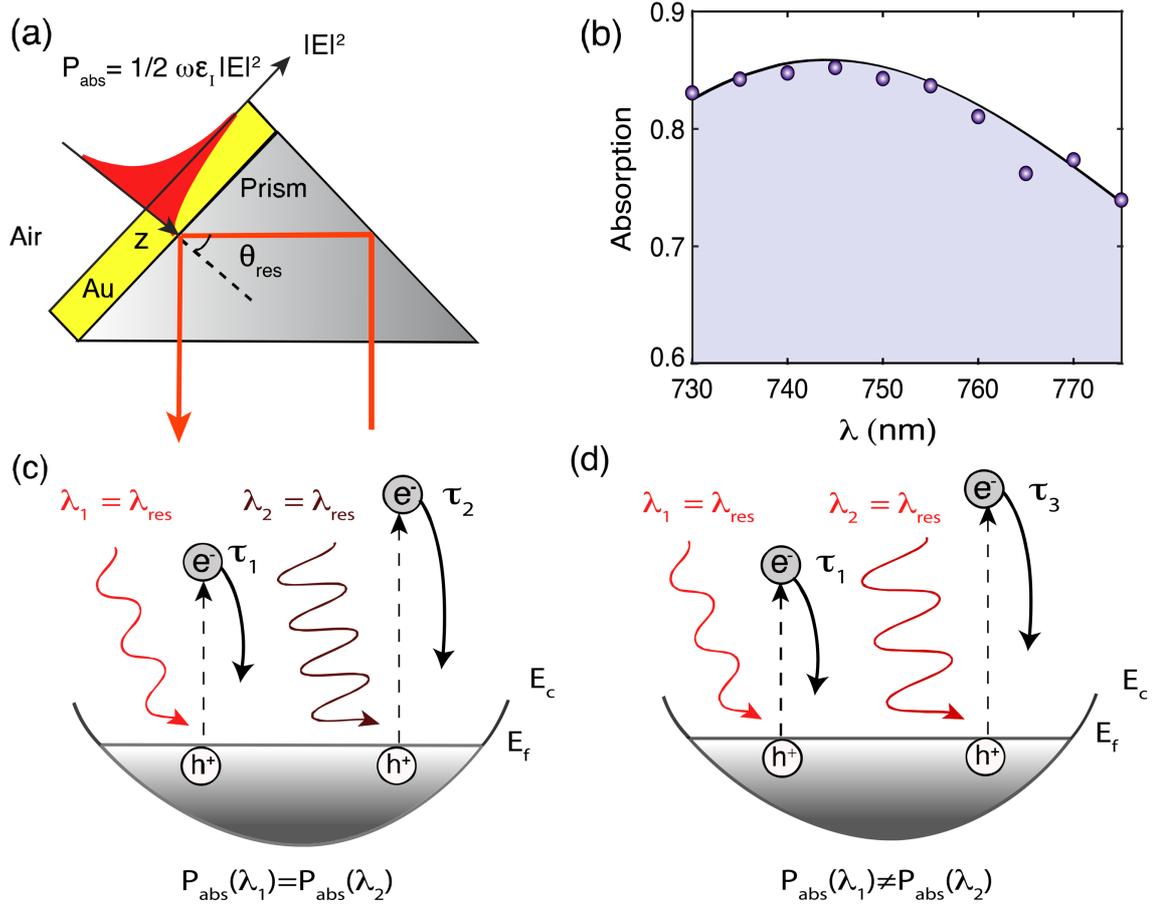

Figure 1: (a) Schematic of light coupling to propagating surface plasmons using the Kretschmann configuration. (b) Absorption measurement (circles) and simulation (solid line) after surface plasmon coupling. (c) Schematic diagram showing hot-electron excitation under resonance and off-resonance wavelengths while keeping the absorbed power fixed (120mW) for both illuminations. (d) Schematic diagram showing a second case where the hot-electron excitation occurs under the same resonance wavelength (745nm) but with different absorbed powers. $\tau_1$, $\tau_2$ and $\tau_3$ are the corresponding electron-phonon relaxation time for these different cases.

Transient reflectivity ($\Delta R/R_0$) measurements as a function of time delay ($\Delta t$) between the pump and probe for both conditions are shown in Figure 2. When the wavelength was varied, we adjust the incident pump intensity according to the absorption spectra (Fig. 1b) to ensure that the absorbed power remains the same over the entire incident wavelength range. We observe that the transient reflectivity ($\Delta R/R_0$) reaches the maximum at resonance, and signal modulation is gradually reduced as the wavelengths tend away from resonance. For the case of fixed wavelength illumination, the input power is varied (59 mW, 105 mW, 141 mW and 176 mW) at the resonance wavelength.

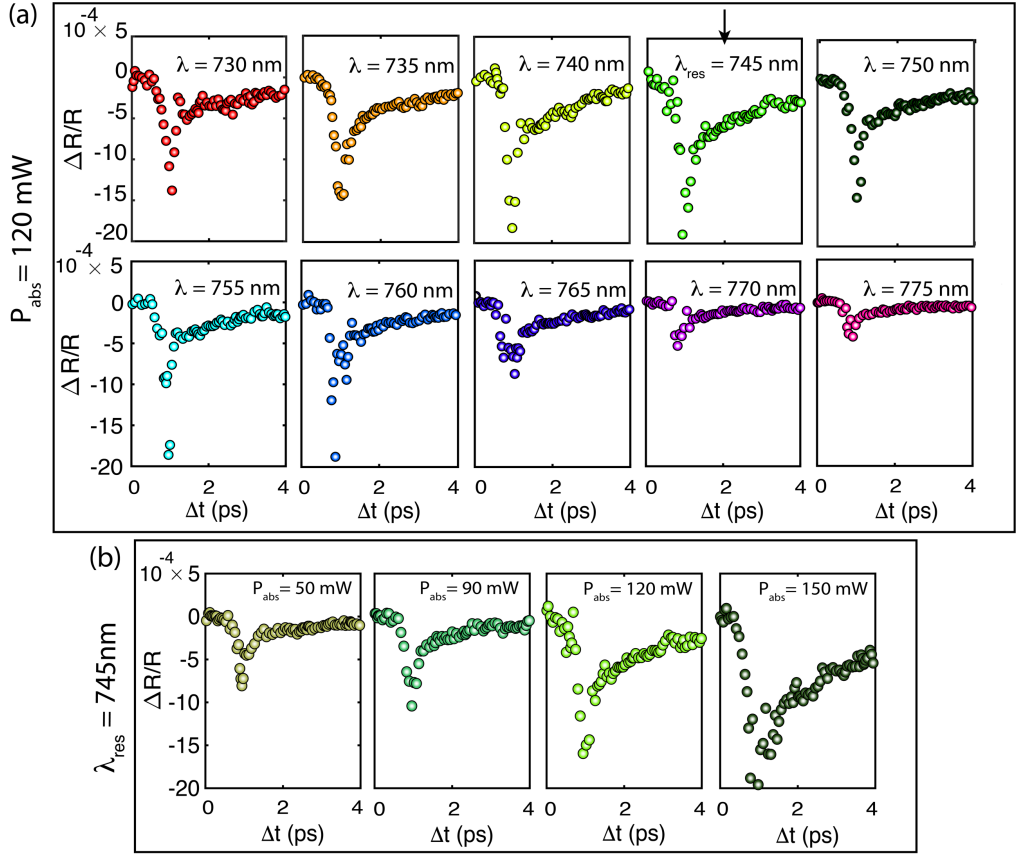

Figure 2: (a) Relative reflectivity change for different incident wavelengths ranging from 730 nm to 775 nm measured at fixed absorbed power (120 mW). Resonance wavelength is distinguished by a green frame from the rest of the wavelengths. (b) Relative reflectivity signals under the fixed 745 nm resonance wavelength measured with the different absorbed powers (50 mW, 90 mW, 120 mW, 150mW).

Transient reflectivity ($\Delta R/R_0$) can be converted to the electron temperature under the intra-band optical pumping, which results in a non-equilibrium hot electron distribution that can modify the optical properties of the Au film. Au band structure is modelled using a simplified parabolic electron density of states[42]. Considering that the carrier density is a temperature independent quantity and the intra-band excitation does not generate extra carriers in the conduction band ($N_{e\_pump} = N_{e\_no\ pump} = 5.049 \times 10^{22}$ cm$^{-3}$ ), we can calculate the chemical potential, Drude plasma frequency ($\omega_p = \sqrt{\frac{e^2 N_e}{\varepsilon_0 \varepsilon_\infty m_e^* m_0}}$) and damping coefficient ($\Gamma_p = \frac{\hbar e}{m_e^* m_0 \mu_e}$) as a function of the electron temperature, where $N_e$ is the carrier concentration, $m_0$ is the mass of electron, $m_e^*$ is the effective mass of electron, and $\mu_e$ is the electron mobility (see the supplementary information for the free electron model details). Subsequently, the change in the reflectivity with electron temperature over different incident wavelengths can be determined from the Transfer Matrix Methods (TMM) calculation, as shown in Figure 3a. Typically, in order to extract the relaxation time of a nonequilibrium system, direct fitting of the TTM is applied to the transient spectroscopic measurements[38,43,44]. However, it is noted that the change in the reflectivity is not linearly proportional to the electron temperature, hence for clear comparison and to better estimate the hot carrier relaxation dynamics, the direct fitting of the TTM is performed on the electron temperature extracted from the transient reflectivity measurements.

Figure 3b and 3c show the converted electron temperature as a function of time delay for both fixed absorbed power with varied wavelengths, and for fixed resonance wavelength with varied absorbed powers.

The converted electron temperatures can be modelled using the TTM as a function of depth from surface (z) and time (t) as follows:

$$C_e(T_e)\left(\frac{\partial T_e}{\partial t}\right) = K_e \nabla^2 T_e - G(T_e - T_l) + S(z,t)$$

$$C_l\left(\frac{\partial T_l}{\partial t}\right) = G(T_e - T_l)$$

where $T_e$ and $T_l$ are electron and lattice temperature [45,46], $C_e(T_e) = \frac{\pi^2 N_e}{2} k_b \left(\frac{k_b T_e}{E_f}\right)$ and $C_l = 2.5 \times 10^6 \, \text{Jm}^{-3}\text{K}^{-1}$ are the electron and lattice heat capacities[45,47], $E_f$ and $k_b$ are the Fermi level and Boltzmann constant, $K_e = 315 \, \text{Wm}^{-1}\text{K}^{-1}$ is the electron thermal conductivity, $G = \frac{C_e(T_e)}{\tau_{e-ph}}$ is the electron-phonon coupling coefficient within the weak perturbation approximation with $\tau_{e-ph}$ as the electron-phonon relaxation time. In general, the skin depth of a material is simply applied to the laser heating source term ($S(z,t)$) to model the laser interaction with the material as a function of depth. Here, we modify the source term ($S(z,t)$) by using the decaying length of the confined electric field of the surface plasmon at both sides of the interface instead of skin depth of the Au (see Fig. S4 for the electrical field profile of Au thin film). The absorbed profiles are calculated from the Finite Difference Time Domain (FDTD) simulations. The calculated field is fitted with double exponential terms, including the decaying field at the Au/prism interface and the decaying field at the Au/air interface. The modified source term ($S(z,t)$) to incorporate the absorbed power profile inside the Au film can be described as:

$$S(z,t) = \sqrt{\frac{\beta}{\pi}} \frac{(1-R)\phi}{t_p} \left( \frac{a_1}{b_1} e^{-\frac{z}{b_1}} + \frac{a_2}{b_2} e^{\frac{z-d}{b_2}} \right) e^{-\beta \left( \frac{t-2t_p}{t_p} \right)^2}$$

where $t_p$ is the laser pulse width, $\phi$ is the laser fluence, d is the sample thickness and $\beta = 4\ln(2)$ [46]. $a_1$ and $a_2$ correspond to the intensity of electric field at Au/air and Au/prism, and $b_1$ and $b_2$ correspond to the decaying length of electric field at Au/air and Au/prism, respectively.

Using our experimental conditions with the modified TTM, we numerically calculate the electron temperature as displayed in Fig. 3b and 3c. For the case of constant absorbed power, we show four wavelengths and their corresponding best fits on the relaxation time to preserve space. The complete set of ten wavelengths are presented in Fig. S2. We also incorporate the spatial dependence of the electron temperature by averaging the temperature profiles along the $z$ direction. The result of the fits is shown in Fig. 3b and 3c based on the Normalized Minimum Squared Error (NMSE) calculation for the hot electron relaxation time. The good agreement of the calculated maximum temperature by TTM and the converted maximum temperature by the free electron model indicates that our free electron model is described well by electron temperature, because the calculated temperature using the TTM solely depends on the experimental conditions.

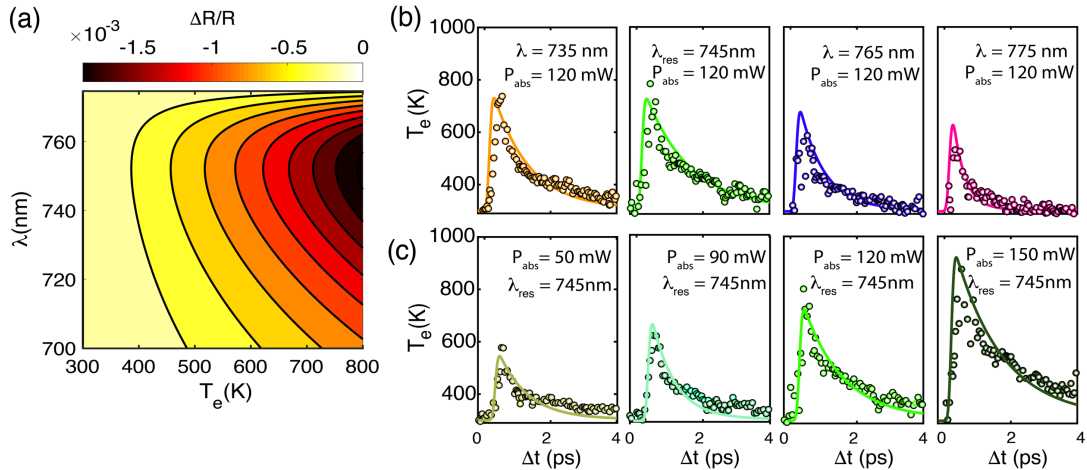

Figure 3: (a) Differential reflectivity contour plot computed from the free electron model and transfer matrix methods. Hot-electron temperature as a function of the delay time between pump and probe beams under (b) fixed (120 mW) absorbed power and (c) fixed resonance wavelength (745 nm). The solid lines are the calculated electron temperatures, and the open circles are the electron temperatures obtained from our differential reflectivity measurements.

Figure 4a and 4b present the extracted hot carrier relaxation time for both cases of fixed absorbed power and fixed illumination wavelength. When the incident power is varied while coupling to the surface plasmon (Fig. 4b), the hot carrier relaxation time increases linearly with increasing incident pump power (Fig. 4b, d). However, when the absorbed power is held constant and the internal field intensity profile is varied (i.e. the amount of surface plasmon coupling is varied), we find that the hot carrier relaxation time is strongly dependent on the intensity of the electric field (see the trend of hot carrier relaxation time in Fig.4a and the normalized maximum intensity of electric field in Fig. 4c). This result confirms that the surface plasmon coupling can enhance the hot carrier relaxation time in the Au film with high field confinement as well as the increase of the light absorption in the Au film. Notably, we can more effectively increase the hot carrier relaxation time with the local electric field enhancement than with increasing the input power. We achieve approximately a doubling of the hot carrier relaxation time with only a ~3.5% increase in electric field intensity (normalized to the input field) at the metal/air interface through SP coupling. Consequently, this feature suggests that electric field confinement helps to excite free electrons to higher energy states, and these non-equilibrium hot electrons take longer to relax via a series of electron-phonon scattering processes.

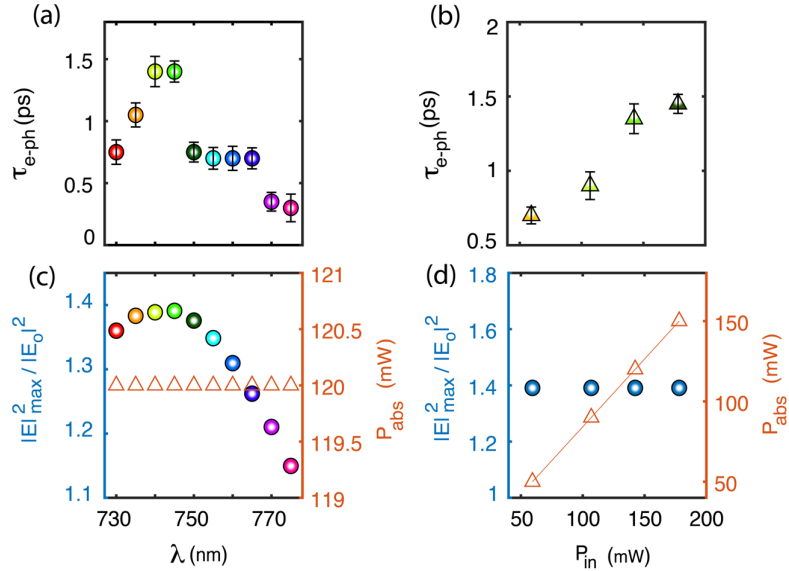

Figure 4: Effect of field enhancement on relaxation time due to the surface plasmon coupling under the fixed (120mW) and variable (50 mW, 90 mW, 120 mW, 150 W) absorbed powers. Experimentally measured hot-electron relaxation time under (a) fixed and (b) variable absorbed powers. Field enhancement computed from the FDTD simulation for wavelengths ranging from 730 nm to 775 nm under the (c) fixed and (d) variable absorbed powers. The electric field profiles are normalized by the intensity of the input field.

**Conclusion**

In summary, we have experimentally demonstrated the impact of propagating surface plasmon excitation on the hot carrier relaxation time through the use of a degenerate pump-probe technique under the Kretschmann configuration. We introduce an approach to analyse the unique internal field confinement in Au thin films with surface plasmon coupling by modifying the two-temperature model. From the comparison study between the constant absorbed pump power and the constant electric field, we determine that the electric field confinement results in the generation of long-lived hot electrons in the Au thin film. Our results provide a foundation for the design of efficient plasmonic systems to tailor hot carrier lifetime with low power consumption in hot carrier based optoelectronic devices.

**Methods**

**Sample fabrication**

A 10 mm N-BK7 right angle prism with an AR coating (wavelength range 650 – 1050 nm) on the face of the hypotenuse is used for the prism coupling. The prism is first cleaned using acetone, methanol, and isopropyl alcohol and then is dried with $N_2$ gas before any deposition processes. An e-beam evaporator is used for the Au deposition at a starting pressure of $3 \times 10^{-6}$ Torr with $0.2$ nm s$^{-1}$ deposition rate. A sample microscope slide is also mounted under the same vacuum chamber and deposition condition for further optical characterization.

**Ellipsometry measurement**

A Wollam-M2000 spectroscopic ellipsometer in the horizontal mode is used for the optical characterization of the sample deposited on a microscope slide. Ellipsometric data is fitted using 2 Lorentz and 1 Drude terms. The results of the permittivity data from the fits are then used for the optical simulations.

**Absorption measurement**

We use a precise motorized rotational mount with 25 arcsecond angular resolution for coupling to the propagating surface plasmon. The incident beam from the glass interface is focused on the Au side of the prism using the off-axis parabolic mirror. Both reflection and transmission of the incoming beam are recorded while rotating the prism automatically. Transmission of the sample is measured to be less than 1% and therefore is negligible for determination of the absorption ($A = 1 - R$). To incorporate possible scattering effects from every interface of the prism, we optimize our absorption measurement using a bare prism first, without any Au coating, to measure the baseline of the reflection signal. The bare prism is then replaced by the Au-coated prism on the rotational stage for the surface plasmon coupling. The reflection signal is recorded over the incident angle for the different pump wavelengths. The final signal is the ratio between the reflectivity measured using the coated and uncoated prisms.

**Transient differential reflectivity measurement**

For the time-resolved differential reflectivity measurements, we employ a degenerate pump-probe technique. Transverse-magnetic (TM) polarized pulses are produced from a femtosecond high power Ti-Sapphire laser with 80 MHZ repetition rate. Using a polarized beam splitter, the incoming pulses are then separated into pump and probe paths. Both beams are directed to coincide on the gold surface after reflecting off the off-axis parabolic mirror to a spot size of approximately 40 $\mu$m. To optimize the signal, the overlap between the two beams is monitored using an AmScope MU1000 digital microscope camera. After spatially separating the two beams, the probe beam is then directed to the Si photodetector for differential reflectivity measurements. The time delay between the pump and probe pulses are produced by passing the pump beam through the mechanical delay stage.

**Numerical Simulations**

The commercially available software (Lumerical FDTD Solutions) is used for the Finite Difference Time Domain (FDTD) calculations. The customized Gaussian source with 150 fs pulse width is generated for the different illumination wavelengths. To avoid extra reflections, Perfectly Matched Layers (PML) are used as the FDTD boundary regions. The electric field profile is simulated within the sample under the varied input powers for the total constant absorbed power throughout the range of the wavelengths.

**Contributions:**

S.M, J.K, J.N.M, and T.E.M developed the experimental concept. S.M carried out the sample preparation and linear characterization. S.M and Y.A performed the pump-probe characterization. S.M and J.K developed the numerical simulations and two-temperature model for the nonlinear response of Au thin film. J.N.M conceived and oversaw the project, and T.E.M provided the facilities. All the authors contributed and approved the manuscript.

**Acknowledgements**

This material is based upon work supported by the National Science Foundation CAREER Grant No. ECCS-1554503 and the Office of Naval Research YIP Award under Grant No. N00014-16-1-2540. The authors also thank Lisa Krayer and Tristan Deppe for their helpful suggestions, comments, and discussions.

**Competing financial interests:**

The authors declare no competing financial interests.

**Author Information:**

†These authors contributed equally to the manuscripts.

*E-mail: jnmunday@ucdavis.edu